PSU-TH/242

# Uses of a Quantum Master Inequality

By

Gordon N. Fleming*

104 Davey Lab
The Pennsylvania State University
University Park, PA 16802

**abstract**

An inequality in quantum mechanics, which does not appear to be well known, is derived by elementary means and shown to be quite useful. The inequality applies to 'all' operators and 'all' pairs of quantum states, including mixed states. It generalizes the rule of the orthogonality of eigenvectors for distinct eigenvalues and is shown to imply all the Robertson generalized uncertainty relations. It severely constrains the difference between probabilities obtained from 'close' quantum states and the different responses they can have to unitary transformations. Thus, it is dubbed a master inequality. With appropriate definitions the inequality also holds throughout general probability theory and appears not to be well known there either. That classical inequality is obtained here in an appendix. The quantum inequality can be obtained from the classical version but a more direct quantum approach is employed here. A similar but weaker classical inequality has been reported by Uffink and van Lith.

* gnf1@earthlink.net

## 1. Introduction

This note is an advertisement, primarily for an inequality and secondarily for the method of derivation. It provides an elementary derivation of and suggests useful employment for a simple but powerful inequality which the author has not seen before and which inquiries have suggested is not well known. Uffink and van Lith (1999) obtained a similar but slightly weaker result in classical probability theory which they applied to the derivation of Thermodynamic uncertainty relations. Accordingly, neither the inequality presented here nor the methods employed to obtain it are peculiar to quantum theory. Rather, essentially identical versions of both apply throughout probability theory and these versions are presented here in Appendix I. Some of the ideas involved here are related to the discussions of physical applications of statistical estimation theory by Uffink and collaborators, as subsequent references will indicate. While there are circumstances in which the inequality derived here can be said to be 40% stronger than the quantum version of the Uffink-van Lith inequality, there are also circumstances in which the two inequalities are equivalent. Indeed some of the consequences obtained here from the present inequality could also be obtained from a quantum version of the Uffink-van Lith inequality. A significant difference concerning the two inequalities is that the derivation given here is noticeably simpler and more elementary than that of Uffink and van Lith.

The special interest of the inequality in the quantum context springs from a quantum state being a repository of a vast array of probability distributions. I call the inequality a master inequality because of the great variety of consequences that follow from it, in particular other inequalities over which it supervenes. Most of the consequences displayed here are, themselves, well known and are rederived here to display that variety and to establish the logical status of the master inequality. But some of the consequences (see the last half of § **3** and § **6** ) I have not seen before.

For any two quantum states of a system and any operator defined on the two states, the inequality asserts a uniform upper bound for the ratio of the absolute difference of the expectation values to the sum of the rms deviations of the operator in the two states. For the special case of pure states, $\Psi$ and $\Psi'$, represented by unit norm vectors, $|\psi>$, and $|\psi'>$, respectively, and a self-adjoint operator, $\hat{A}$, the exact relationship is that if,

$$|\langle \psi' | \psi \rangle| := \cos\theta, \qquad 0 \le \theta \le (\pi/2), \tag{1.1}$$

<u>then</u>,

$$\frac{|\langle A \rangle_{\psi'} - \langle A \rangle_{\psi}|}{\Delta_{\psi'}A + \Delta_{\psi}A} \le \tan\theta\ , \tag{1.2}$$

where $\langle A \rangle_{\psi}$ denotes the expectation value and $\Delta_{\psi}A$ the rms deviation of A in the state $\Psi$.

The fact that the same upper bound, tan θ, holds for <u>any</u> self adjoint operator will be used (see § **3**) to give a precise and strong form to the not surprising notion that if tan θ is small, it will be difficult to distinguish between the two states by performing measurements on systems in those states (Hilgevoord and Uffink, 1990, 1991). This, coupled with the form of the bounded ratio, makes the inequality reminiscent of the Rayleigh criterion for distinguishing between optical images of nearby point sources by comparing the separation of their diffraction peaks with the widths of the diffraction circles. But uses of (1.2) will also be indicated for the case in which tan θ is not small.

By inspection we see that (1.1,2) implies the rule that eigenvectors for distinct eigenvalues of self adjoint operators must be orthogonal. In fact the inequality generalizes that rule by forcing the approach to orthogonality as the rms deviations decrease in the presence of fixed separation of expectation values or as the separation increases in the presence of fixed rms deviations. In § **3** we will see explicit instances of this forcing.

In § **4** it will be shown that all of the Robertson (1929) generalized uncertainty relations,

$$\Delta_{\psi}A\, \Delta_{\psi}B \ge (1/2)|\langle [\hat{A}, \hat{B}] \rangle_{\psi}|, \tag{1.3}$$

emerge as consequences of special cases of the inequality. The generalization to non-self adjoint operators and mixed quantum states is presented in § **5** and the example of unitary operators is considered in § **6**.

## 2. The basic derivation

For any unit norm state vector, $|\psi>$, and any self adjoint operator, $\hat{A}$, define the vector, $|\psi_A>$, by

$$|\psi_A> := \frac{(\hat{A} - <A>_\psi)|\psi>}{\Delta_\psi A} , \tag{2.1}$$

where, $<A>_\psi := <\psi|\hat{A}|\psi>$, and $(\Delta_\psi A)^2 := <\psi|(\hat{A} - <A>_\psi)^2|\psi>$, and both quantities are assumed finite. Then

$$<\psi|\psi_A> = 0, \qquad <\psi_A|\psi_A> = 1, \tag{2.2}$$

and

$$\hat{A}|\psi> = |\psi><A>_\psi + |\psi_A>\Delta_\psi A . \tag{2.3}$$

Next consider, $<\psi'|\hat{A}|\psi>$. Applying (2.3) in both directions we have,

$$<\psi'|\hat{A}|\psi> = <\psi'|\psi><A>_\psi + <\psi'|\psi_A>\Delta_\psi A$$

$$= <A>_{\psi'}<\psi'|\psi> + \Delta_{\psi'}A<\psi'_A|\psi> , \tag{2.4}$$

or

$$(<A>_{\psi'} - <A>_\psi)<\psi'|\psi>$$

$$= <\psi'|\psi_A>\Delta_\psi A - \Delta_{\psi'}A<\psi'_A|\psi> . \tag{2.5}$$

Hence, taking absolute values,

$$|<A>_{\psi'} - <A>_\psi||<\psi'|\psi>|$$

$$\leq \ (|<\psi'|\psi_A>|\Delta_\psi A + \Delta_{\psi'}A|<\psi'_A|\psi>|) . \tag{2.6}$$

Setting

$$|\langle \psi ' | \psi \rangle| := \cos\theta , \qquad 0 \le \theta \le \pi/2 , \tag{2.7}$$

then from,

$$\| |\psi\rangle - |\psi '\rangle\langle \psi ' | \psi\rangle - |\psi '_A\rangle\langle \psi '_A | \psi\rangle \|^2 \ \ge \ 0 \tag{2.8a}$$

and, $$\| |\psi '\rangle - |\psi\rangle\langle \psi | \psi '\rangle - |\psi_A\rangle\langle \psi_A | \psi '\rangle \|^2 \ \ge \ 0 \tag{2.8b}$$

or, equivalently,

$$1 = \langle \psi | \psi \rangle \ \ge \ |\langle \psi ' | \psi \rangle|^2 + |\langle \psi '_A | \psi \rangle|^2 \ , \tag{2.9a}$$

and, $$1 = \langle \psi ' | \psi ' \rangle \ \ge \ |\langle \psi ' | \psi \rangle|^2 + |\langle \psi ' | \psi_A \rangle|^2 \ , \tag{2.9b}$$

we have,

$$|\langle \psi '_A | \psi \rangle| , |\langle \psi ' | \psi_A \rangle| \ \le \ \sin\theta \ . \tag{2.10}$$

Substituting (2.7) and (2.10) into (2.6) we find,

$$| \langle A \rangle_{\psi'} - \langle A \rangle_{\psi} | \cos\theta \ \le \ ( \Delta_{\psi'} A + \Delta_{\psi} A ) \sin\theta , \tag{2.11}$$

or, the desired inequality, (1.2).

## 3. Some applications

From the monotonically increasing character of $\tan\theta$ and the monotonically decreasing character of $\cos\theta$, in the given range, it follows that if, for some given self adjoint $\hat{A}$, we have

$$\frac{| \langle A \rangle_{\psi'} - \langle A \rangle_{\psi} |}{\Delta_{\psi'} A + \Delta_{\psi} A} \ = \ \tan\theta \ , \tag{3.1}$$

then $$| \langle \psi ' | \psi \rangle | \ \le \ \cos\theta \ . \tag{3.2}$$

This generalizes the familiar rule that eigenvectors ( $\Delta_{\psi}A = \Delta_{\psi'}A = 0$ ) corresponding to distinct eigenvalues ( $<A>_{\psi'} \neq <A>_{\psi}$ ) are orthogonal ( $\tan\theta = \infty$ and $\cos\theta = 0$ ).

For example, let $\hat{X}$ be a position operator, $\hat{P}$ the canonically conjugate momentum, and let

$$|\psi'> = \exp[(i/\hbar)\,\hat{P}\,d]\,|\,\psi>. \qquad (3.3)$$

Then,

$$\frac{|<X>_{\psi'} - <X>_{\psi}|}{\Delta_{\psi'}X + \Delta_{\psi}X} = \frac{d}{2\Delta_{\psi}X} = \tan\theta\ , \qquad (3.4)$$

yields, (L'evy-Leblond, 1985), (Uffink, 1994),

$$|<\psi\,|\exp[(i/\hbar)\,\hat{P}\,d]\,|\,\psi>| \ \leq\ \cos\theta = \frac{2\Delta_{\psi}X}{\sqrt{d^2 + (2\Delta_{\psi}X)^2}}, \qquad (3.5)$$

an upper bound for the inner product between two relatively spatially displaced states. By interchanging the roles of $\hat{X}$ and $\hat{P}$ we similarly obtain

$$|<\psi\,|\exp[(i/\hbar)\,\hat{X}\,q]\,|\,\psi>| \ \leq\ \cos\theta = \frac{2\Delta_{\psi}P}{\sqrt{q^2 + (2\Delta_{\psi}P)^2}}\ . \qquad (3.6)$$

If $\hat{X}$ is a center of mass position coordinate and we write $q = M\,v$ , where $M$ is the total mass of the system, then (3.6) yields an upper bound for the inner product between two relatively Galilean boosted states.

Returning to (1.2) consider the case in which $\hat{A} = \hat{\Pi}$ , a projection operator, i.e.,

$$\frac{|<\Pi>_{\psi'} - <\Pi>_{\psi}|}{\Delta_{\psi'}\Pi + \Delta_{\psi}\Pi} \leq \tan\theta\,. \qquad (3.7a)$$

The importance of this case lies in the fact that <u>any</u> probability that can be obtained from a quantum state is expressible as the expectation value of some projection operator. Therefore, for $\psi$ and $\psi'$ there is no probability that differs between them by more than (3.7a) allows .

Furthermore, for any projection operator, the rms deviation is,

$$\Delta_\psi \Pi = \sqrt{<\Pi>_\psi (1 - <\Pi>_\psi)} \ , \tag{3.7b}$$

an explicit function of the expectation value. This function always lies between 0 and 1/2 and approaches zero for both the zero and unit limit of the expectation value. These features of the rms deviation make (3.7a) a rather powerful inequality.

To see this, recall that regardless of how large $\tan\theta$ might be, so long as it is finite, the absolute difference of expectation values can not equal unity. For the difference to equal unity, the two expectation values would have to be zero and unity, in which case they would be eigenvalues and the two states, eigenstates corresponding to distinct eigenvalues, i.e., the states would be orthogonal and $\tan\theta$, infinite. Now the inequality, (3.7a), enforces this result due to (3.7b), i.e., for any finite $\tan\theta$ , it places limits below unity on the absolute difference of expectation values. In fact, a variational analysis of (3.7a,b), which we provide in Appendix II, yields the useful, but weaker inequality,

$$|<\Pi>_{\psi'} - <\Pi>_\psi| \le \sin\theta . \tag{3.8a}$$

To exemplify the weakness of (3.8a), relative to (3.7a,b), we mention that the latter also directly implies that if one of the expectation values is zero or unity, then,

$$|<\Pi>_{\psi'} - <\Pi>_\psi| \le (\sin\theta)^2 . \tag{3.8b}$$

These last two inequalities are especially provocative at the other extreme of small values of $\theta$. Here they provide a precise sense in which it would be difficult to distinguish statistically between $\psi$ and $\psi'$. The smaller the value of $\theta$, the finer the statistical data would have to be to make the distinction between the probabilities represented by the expectation values of the projection operator. An illuminating discussion of this kind of problem in the context of general probability theory is provided by Hilgevoord and Uffink, (1991). An earlier discussion by the same authors focused on quantum mechanics is (Hilgevoord and Uffink, 1990).

Taking our next departure from an aspect of the discussion of these authors, we consider the case in which the state vectors $|\psi>$ and $|\psi'>$ are two members of a family, differentiably parameterized by the continuous variable, b, i.e.,

$$|\psi>=|\psi(b)> \qquad \text{and} \qquad |\psi'>=|\psi(b+\delta b)>. \tag{3.9}$$

Then, if $\delta b$ is infinitesimal, and keeping in mind the unit norm of the state vectors, we have,

$$|<\psi(b)|\psi(b+\delta b)>|=1-\frac{\Delta[\psi,b]^2}{2}\delta b^2+O(\delta b^4)\ , \tag{3.10a}$$

where

$$\Delta[\psi,b]:=\sqrt{\|\frac{d}{db}|\psi(b)>\|^2-|<\psi(b)|\frac{d}{db}|\psi(b)>|^2}\ \geq\ 0. \tag{3.10b}$$

If we now write,

$$<\Pi>_{\psi(b)}:=(\cos\phi(b))^2, \tag{3.11}$$

then, to lowest order in $\delta b$,

$$|<\Pi>_{\psi(b+\delta b)}-<\Pi>_{\psi(b)}|=2|\cos\phi(b)\sin\phi(b)\frac{d}{db}\phi(b)\,\delta b|, \tag{3.12a}$$

$$(\Delta_{\psi(b+\delta b)}\Pi+\Delta_{\psi(b)}\Pi)=2|\cos\phi(b)\sin\phi(b)|\ , \tag{3.12b}$$

and

$$\tan\theta=\Delta[\psi,b]\,\delta b\,. \tag{3.12c}$$

Consequently, (3.7a) implies,

$$|\frac{d}{db}\phi(b)|\leq\Delta[\psi,b] \quad \text{for} \quad \forall b. \tag{3.13}$$

An interesting example of (3.13) is provided by letting $|\psi(b)>$ be given by a unitary transformation,

$$|\psi(b)> := \exp[i\,\hat{B}\,b]\,|\psi>, \tag{3.14}$$

with self adjoint $\hat{B}$. From (3.10b) we then have,

$$\Delta[\psi, b] = \Delta_\psi B\ . \tag{3.15}$$

For $0 \leq \Delta_\psi B\, b \leq (\pi/2)$, (3.11, 13, 15) yield,

$$|<\Pi>_{\psi(b)}| \geq (\cos(\Delta_\psi B\, b))^2. \tag{3.16}$$

As is nicely spelled out by Uffink (1993), relations of this type for $b$ = time and $B$ = energy and,

$$\hat{\Pi} = |\psi><\psi|, \tag{3.17a}$$

with

$$<\Pi>_{\psi(b)} = |<\psi|\exp[\,i\,\hat{B}\,b]\,|\psi>|^2, \tag{3.17b}$$

were originally obtained by Mandelstam and Tamm, (1945), and then rederived by several workers, (Fleming, 1973), (Bhattacharya, 1983), (Home and Whitaker, 1986), (Vaidman, 1992). The case of arbitrary 'canonical' pairs, (b, B), is presented, amidst other generalizations, in (Fleming, 1981). The interest here lies in the emergence of (3.13-17) from special cases of the inequality (3.7a) or (1.2).

Returning to (3.5) we apply (3.16, 17), with $\hat{B}$ replaced by $\hat{P}$ and $b$ replaced by $d$, to find that, for $0 \leq \frac{\Delta_\psi P}{\hbar} d \leq \frac{\pi}{2}$, we have,

$$\cos\left(\frac{\Delta_\psi P}{\hbar} d\right) \leq |<\psi|\exp[(i/\hbar)\,\hat{P}\,d]\,|\psi>|\ , \tag{3.18}$$

which complements (3.5). Similarly, (3.6) is complemented by

$$\cos\left(\frac{\Delta_\psi X}{\hbar} q\right) \;\le\; |\langle \psi \,|\, \exp[(i/\hbar)\,\hat{X}\, q]\,|\, \psi \rangle| \;\; . \tag{3.19}$$

As a precursor to the next section we note that the pair, (3.5, 18), <u>or</u> the pair, (3.6, 19), yield Heisenberg's uncertainty relation.

## 4. The Robertson generalized uncertainty relations

To obtain, from the inequality (1.2), the Robertson generalized uncertainty relations (1.3) (Robertson, 1929), we simply choose,

$$|\psi'\rangle := \exp[i\,\hat{B}\,\delta b]\,|\,\psi\rangle$$

$$= (1 + i\,\hat{B}\,\delta b - (1/2)\,\hat{B}^2\,\delta b^2\,)\,|\,\psi\rangle \;+\; O(\delta b^3)\;, \tag{4.1}$$

or

$$\cos\,\theta \;=\; |\langle \psi' | \psi \rangle| \;=\; 1 - (1/2)\,(\Delta_\psi B\,\delta b)^2 + O(\delta b^4)\;. \tag{4.2}$$

This yields,

$$|\langle A \rangle_{\psi'} - \langle A \rangle_\psi| \;=\; |\langle [\,\hat{A}\,,\hat{B}\,]\rangle_\psi\;\delta b| + O(\delta b^2)\;, \tag{4.3}$$

$$\Delta_{\psi'} A + \Delta_\psi A = 2\Delta_\psi A + O(\delta b), \tag{4.4}$$

and
$$\Delta_\psi B\,|\,\delta b\,| + O(\delta b^2) = \;\tan\theta \;\approx\; \theta . \tag{4.5}$$

Combining these in (1.2) we obtain,

$$\frac{|\langle [\,\hat{A}\,,\hat{B}\,]\rangle_\psi\;\delta b\,|}{2\Delta_\psi A} \;\le\; \Delta_\psi B\,|\,\delta b\,|\;, \tag{4.6}$$

from which the desired relations, (1.3), follow immediately.

## 5. The general case

We now consider the generalization of (1.2) to the case of arbitrary quantum states and operators. First we define the rms deviation for a non-self adjoint operator , $\hat{A}$, in the quantum state represented by the density operator, $\hat{\rho}$, as the non-negative quantity, $\Delta_\rho A$, satisfying

$$(\Delta_\rho A)^2 := <(\hat{A}^\dagger - <A>_\rho{}^*)(\hat{A} - <A>_\rho)>_\rho \ ,$$
(5.1)

where, $$<\hat{A}>_\rho := \mathrm{tr}[\hat{\rho}\hat{A}] \ . \tag{5.2}$$

Notice that notwithstanding the real non-negative character of $\Delta_\rho A$, there can be a difference between $\Delta_\rho A$ and $\Delta_\rho A^\dagger$ for a non-normal operator.

Next we introduce, $\hat{q}$ , as the non-negative operator satisfying,

$$\hat{q}^2 = \hat{\rho} \ . \tag{5.3}$$

Then, when (5.1,2) define finite quantities, we introduce, $\hat{q}_A$ , by

$$\hat{q}_A := \frac{(\hat{A} - <A>_\rho)\hat{q}}{\Delta_\rho A} \ , \tag{5.4}$$

and find, $$\mathrm{tr}[\hat{q}\,\hat{q}_A] = 0 \ , \tag{5.5}$$

and $$\mathrm{tr}[\hat{q}_A{}^\dagger\,\hat{q}_A] = \mathrm{tr}[\hat{\rho}] = 1 . \tag{5.6}$$

From these relations it follows that,

$$\hat{q}'\hat{A}\,\hat{q} \ = \ \hat{q}'\hat{q} <A>_\rho + \ \hat{q}'\hat{q}_A\ \Delta_\rho A$$

$$= <A>_{\rho'}\ \hat{q}'\hat{q}\ + \Delta_{\rho'}A^\dagger\ (\hat{q}'_{A^\dagger})^\dagger\ \hat{q} \ , \tag{5.7}$$

or , upon taking traces and rearranging terms,

$$(< A >_{\rho'} - < A >_{\rho}) \operatorname{tr}[\hat{q}'\hat{q}]$$

$$= \operatorname{tr}[\hat{q}'\hat{q}_A]\Delta_\rho A - \Delta_{\rho'}A^\dagger \operatorname{tr}[(\hat{q}'_{A^\dagger})^\dagger \hat{q}] . \qquad (5.8)$$

Considering absolute values of both sides we obtain,

$$|< A >_{\rho'} - < A >_{\rho}||\operatorname{tr}[\hat{q}'\hat{q}]|$$

$$\leq (|\operatorname{tr}[\hat{q}'\hat{q}_A]|\Delta_\rho A + \Delta_{\rho'}A^\dagger |\operatorname{tr}[(\hat{q}'_{A^\dagger})^\dagger \hat{q}]|). \qquad (5.9)$$

Now put,

$$\operatorname{tr}[\hat{q}'\hat{q}] := \cos\theta , \qquad 0 \leq \theta \leq \pi/2 , \qquad (5.10)$$

and note that from (5.5,6) it follows that,

$$1 = \operatorname{tr}[\hat{q}'^2] \geq \operatorname{tr}[|\hat{q}\operatorname{tr}[\hat{q}'\hat{q}] + \hat{q}_A^\dagger \operatorname{tr}[\hat{q}'\hat{q}_A]|^2] ,$$

$$= |\operatorname{tr}[\hat{q}'\hat{q}]|^2 + |\operatorname{tr}[\hat{q}'\hat{q}_A]|^2 , \qquad (5.11)$$

and

$$1 = \operatorname{tr}[\hat{q}^2] \geq \operatorname{tr}[|\hat{q}'\operatorname{tr}[\hat{q}'\hat{q}] + \hat{q}'_{A^\dagger} \operatorname{tr}[(\hat{q}'_{A^\dagger})^\dagger \hat{q}]|^2]$$

$$= |\operatorname{tr}[\hat{q}'\hat{q}]|^2 + |\operatorname{tr}[(\hat{q}'_{A^\dagger})^\dagger \hat{q}]|^2 . \qquad (5.12)$$

Consequently, from (5.10) we have,

$$|\operatorname{tr}[(\hat{q}'_{A^\dagger})^\dagger \hat{q}]| , \ |\operatorname{tr}[\hat{q}'\hat{q}_A]| \leq \sin\theta . \qquad (5.13)$$

Thus (5.9) yields,

$$|< A >_{\rho'} - < A >_{\rho}|\cos\theta \leq (\Delta_\rho A + \Delta_{\rho'}A^\dagger)\sin\theta , \qquad (5.14)$$

or $$\frac{|<A>_{\rho'} - <A>_{\rho}|}{\Delta_{\rho} A + \Delta_{\rho'} A^{\dagger}} \leq \tan\theta \ . \qquad (5.15)$$

But while we have obtained an inequality of the form we seek, it does not quantitatively reproduce our pure state result, (1.2), (for normal operators), when the density operators, $\hat{\rho}$ and $\hat{\rho}'$ satisfy,

$$\hat{\rho} := |\psi><\psi| \ , \qquad \text{and} \qquad \hat{\rho}' := |\psi'><\psi'| \ . \qquad (5.16)$$

In this case the angle $\theta$ defined in (5.10) is larger than the angle $\theta$ defined in (2.7).

To remedy this defect we employ the polar decomposition theorem, according to which the operator, $\hat{q}'\hat{q}$, has the form,

$$\hat{q}'\hat{q} = \hat{V}|\hat{q}'\hat{q}| \quad , \qquad (5.17a)$$

where $|\hat{q}'\hat{q}|$ is the non-negative operator, $(\hat{q}\,\hat{q}'^2\,\hat{q})^{1/2}$, and $\hat{V}$ is a partial isometry from the orthogonal complement of the null space of $\hat{q}'\hat{q}$ onto the orthogonal complement of the null space of $\hat{q}\,\hat{q}'$. The projection operators onto these equi-dimensional orthogonal complements are thus, $\hat{V}^{\dagger}\hat{V}$ and $\hat{V}\,\hat{V}^{\dagger}$, respectively. Consequently,

$$|\hat{q}'\hat{q}| = \hat{V}^{\dagger}\hat{q}'\hat{q}, \qquad (5.17b)$$

and if we define the angle $\theta$ by*

$$\mathrm{tr}[|\hat{q}'\hat{q}|] := \cos\theta \ , \qquad (5.18)$$

compatibility with the definition (2.7) in the case, (5.16), is achieved. It remains to rederive the inequality, (5.15), in the context of the new definition, (5.18).

* I am indebted to Jos Uffink and Guido Bacciagalluppi, independently, for this suggestion.

To that end we return to (5.7) and multiply from the left by $\hat{V}^\dagger$. Then, upon taking traces and rearranging terms and considering absolute values of both sides, we have,

$$|< A >_{\rho'} - < A >_\rho| \, \mathrm{tr}[\, \hat{V}^\dagger \hat{q}' \hat{q} \,]$$

$$\leq \; (|\, \mathrm{tr}[\, \hat{V}^\dagger \hat{q}' \hat{q}_A \,]|\, \Delta_\rho A + \; \Delta_{\rho'} A^\dagger \, |\, \mathrm{tr}[\, \hat{V}^\dagger (\hat{q}'_{A^\dagger})^\dagger \, \hat{q} \,]|) \, . \tag{5.19}$$

Now given (5.17b, 18), we can regain (5.15) from (5.19) if we can show that,

$$|\, \mathrm{tr}[\, \hat{V}^\dagger \hat{q}' \hat{q}_A \,]| \; , \quad |\, \mathrm{tr}[\, \hat{V}^\dagger (\hat{q}'_{A^\dagger})^\dagger \, \hat{q} \,]| \quad \leq \quad \sin\theta \quad . \tag{5.20}$$

But remembering that $\hat{V}^\dagger \hat{V}$ and $\hat{V}\hat{V}^\dagger$ are projection operators, we can obtain (5.20) as follows. First from,

$$\sin^2\theta \; - \; |\, \mathrm{tr}[\, \hat{V}^\dagger \hat{q}' \hat{q}_A \,]|^2 \; = 1 \; - \; \mathrm{tr}[\, \hat{V}^\dagger \hat{q}' \hat{q} \;\,]^2 \; - \; |\, \mathrm{tr}[\, \hat{V}^\dagger \hat{q}' \hat{q}_A \,]|^2$$

$$\geq \; \mathrm{tr}[\, \hat{V}^\dagger \hat{q}'^2 \, \hat{V} \,] \; - \; \mathrm{tr}[\, \hat{V}^\dagger \hat{q}' \hat{q} \;\,]^2 \; - \; |\, \mathrm{tr}[\, \hat{V}^\dagger \hat{q}' \hat{q}_A \,]|^2$$

$$= \; \mathrm{tr}[\, \hat{C}\hat{C}^\dagger] \; \geq \; 0 \, , \tag{5.21a}$$

where,

$$\hat{C} \; := \; \hat{V}^\dagger \hat{q}' \; - \; \hat{q} \, \mathrm{tr}[\, \hat{V}^\dagger \hat{q}' \hat{q} \,] \; - \; (\hat{q}_A)^\dagger \, \mathrm{tr}[\, \hat{V}^\dagger \hat{q}' \hat{q}_A \,] \; , \tag{5.21b}$$

we get the first inequality in (5.20). Next from,

$$\sin^2\theta \; - \; |\, \mathrm{tr}[\, \hat{V}^\dagger (\hat{q}'_{A^\dagger})^\dagger \hat{q} \,]|^2 \; = 1 \; - \; \mathrm{tr}[\, \hat{V}^\dagger \hat{q}' \hat{q} \,]^2 \; - \; |\, \mathrm{tr}[\, \hat{V}^\dagger (\hat{q}'_{A^\dagger})^\dagger \hat{q} \,]|^2$$

$$\geq \quad \mathrm{tr}[\, \hat{V} \, \hat{q}^2 \hat{V}^\dagger] \; - \; \mathrm{tr}[\, \hat{q}' \, \hat{q} \, \hat{V}^\dagger \,]^2 \; - \; |\, \mathrm{tr}[\, (\hat{q}'_{A^\dagger})^\dagger \, \hat{q} \, \hat{V}^\dagger \,]|^2$$

$$= \; \mathrm{tr}[\, \hat{D}^\dagger \hat{D} \,] \; \geq \; 0 \; , \tag{5.22a}$$

where,

$$\hat{D} := \hat{q}\,\hat{V}^{\dagger} - \hat{q}'\,\mathrm{tr}[\hat{q}'\hat{q}\,\hat{V}^{\dagger}] - (\hat{q}'_{A^{\dagger}})\,\mathrm{tr}[(\hat{q}'_{A^{\dagger}})^{\dagger}\,\hat{q}\,\hat{V}^{\dagger}] \quad , \tag{5.22b}$$

we get the second inequality in (5.20). Thus we have (5.15) with (5.18), as desired.

## 6: Unitary operators

The unitary operators share with projection operators the feature of having their rms deviations in any state (as defined in (5.1)) be an explicit function of their expectation values in that state. Thus for

$$\hat{U}^{-1} = \hat{U}^{\dagger}, \tag{6.1}$$

we have

$$\Delta_{\rho} U = \sqrt{1 - |< U >_{\rho}|^{2}} \;. \tag{6.2}$$

Putting $\quad |< U >_{\rho}| = \cos\alpha \quad$ and $\quad |< U >_{\rho'}| = \cos\alpha',$ (6.3)

where , $\quad 0 \le \alpha\,,\alpha' \le \pi/2\,,$ (6.4)

and noting that,

$$||< U >_{\rho'}| - |< U >_{\rho}|| \le |< U >_{\rho'} - < U >_{\rho}|\;, \tag{6.5}$$

it follows from (6.2-5) and (5.16), with U replacing A, that

$$\frac{|\cos\alpha' - \cos\alpha\,|}{\sin\alpha' + \sin\alpha} \le \tan\theta. \tag{6.6}$$

Writing $\alpha'$ and $\alpha$ as

$$\begin{pmatrix}\alpha' \\ \alpha\end{pmatrix} = \left(\frac{\alpha' + \alpha}{2}\right) \pm \left(\frac{\alpha' - \alpha}{2}\right) \quad , \tag{6.7}$$

trigonometric identities, applied to the left side of (6.6), yield,

$$\frac{|\alpha' - \alpha|}{2} \leq \theta \; . \tag{6.8}$$

Considering the defined range of the angles, θ, α, and α', the inequality (6.8) tells us something only for $0 \leq \theta \leq \pi/4$.

For an application of (6.8) let us return to the pure state case and assume again the relations (3.9,10). We then have (3.12c), and recognizing $\alpha$ and $\alpha'$ as functions of the parameter, b, we immediately obtain from (6.8),

$$|\frac{d}{db}\alpha(b)| \leq 2\,\Delta[\psi, b], \quad \text{for} \quad \forall b \; . \tag{6.9}$$

This novel result is the version for expectation values of <u>unitary</u> operators corresponding to the result (3.13) for <u>projection</u> operators. Note that the difference of a factor of 2 between (6.9) and (3.13) correlates with the difference of 2 in the exponent in the definitions, (6.3) and (3.11), of $\cos\alpha$ and, $\cos\phi(b)$ respectively. The inequality, (6.9), generalizes easily to the case of mixed states.

In special cases one would expect to be able to obtain stronger results than (6.8). Indeed, Uffink and Hilgevoord (1985), in the context of distinguishing between the uncertainty <u>principle</u> and the uncertainty <u>relations</u>, have examined a comparison of the expectation values of a unitary operator in two <u>pure</u> states related by a projection operator, i.e.,

$$|\psi'\rangle := \frac{\hat{\Pi}|\psi\rangle}{\sqrt{\langle\psi|\hat{\Pi}|\psi\rangle}} \; , \tag{6.10}$$

where

$$[\hat{\Pi}, \hat{U}] = 0 \, . \tag{6.11}$$

They derive an inequality (not their final result) which, in our notation and modulo our earlier comment about the distinction in the angle $\theta$ for pure states and mixed states, is

$$\cos\alpha' \leq 1 + \cos\alpha - (\cos\theta)^2 \; . \tag{6.12}$$

This is stronger than (6.8) for the range, $\pi/2 \geq \alpha \geq \max(2\theta, \pi/3)$, in which they use it.

Finally, one can sometimes obtain, from (5.16), a limit on the difference of the phases of $< U >_\rho$ and $< U >_{\rho'}$ if one takes their absolute values as given. Calling the absolute values, M, and M', respectively we find from (5.16), with A replaced by U,

$$(1 - \cos \Delta\phi) \leq \frac{1}{2MM'}[ ( \sqrt{1- M'^2} + \sqrt{1 - M^2} )^2 \tan^2\theta - ( M' - M )^2] , \quad (6.13)$$

where $\Delta\phi$ is the difference of the phases. The right hand side has to be less than 2 for (6.13) to provide a limit on $\Delta\phi$.

Furthermore, suppose $\rho$ is an eigenstate of the unitary transformation, e.g. $\hat{U}$ implements a rotation about the z axis while $\rho$ is an eigenstate of the z component of total angular momentum. Then $M = 1$ and (6.13) becomes,

$$(1 - \cos \Delta\phi) \leq \frac{1 - M'}{2M'}[(1 + M') \tan^2\theta - (1 - M')] . \quad (6.14a)$$

Now let the states, $\rho'$ and $\rho$ be 'close' enough to each other so that $\tan \theta \leq 1$. Then, since the quantity inside the square brackets must be non-negative, we have,

$$M' \geq \cos 2\theta, \quad (6.14b)$$

and only if M' exceeds this lower bound, while remaining below the upper bound of unity, is there room for $\Delta\phi$ to be non-zero.

**7: Conclusion and acknowledgement**

I have presented an elementary derivation of an apparently new, simple and powerful inequality in quantum theory. I have tried to illustrate the breadth and strength of the inequality by considering a variety of examples and applications of it, some of them yielding novel results. The logical status of the inequality within quantum mechanics seems to be quite high and it is obvious that only the surface of applications has been scratched in this paper. Further examination of this master inequality seems to be warranted.

I wish to thank Jos Uffink for very helpful communications during the early stages of the preparation of this paper.

**Appendix I:**

Let P and P' be probability distributions over (the set of variables) x with respect to the measure, μ, i.e.

$$\int d\mu(x)\, P(x) \;=\; \int d\mu(x)\, P'(x) \;=\; 1\,. \tag{I.1}$$

Let Q and Q' be the non-negative square roots of P and P', respectively, and let A be a function of x with finite expectation values and rms deviations under the distributions, P and P'. If A is complex, which is permitted, then the rms deviation is defined as,

$$\Delta_P A \;:=\left|\, \sqrt{<|A|^2>_P - |<A>_P|^2}\,\right|, \tag{I.2}$$

and similarly for $\Delta_{P'}A$. Now if we define $\theta$ ( $0 \le \theta \le \pi/2$ ) by,

$$(Q', Q) \;:=\; \cos\theta\;, \tag{I.3}$$

where,

$$(\,F, G\,) \;:= \int d\mu(x)\, F^*(x)\, G(x)\,, \tag{I.4}$$

(the complex conjugation will be needed later), then we claim,

$$\frac{|<A>_{P'} - <A>_P|}{\Delta_{P'}A + \Delta_P A} \;\le\; \tan\theta\;\;. \tag{I.5}$$

To see this, first define the function, $Q_A$, by,

$$Q_A(x) := \frac{1}{\Delta_P A}\,(\,A(x) - <A>_P)\, Q(x)\,. \tag{I.6}$$

and similarly, $Q'_A$. We then have,

$$(Q, Q_A) = 0 = (Q', Q'_A), \text{ and } (Q_A, Q_A) = 1 = (Q_A', Q_A') , \tag{I.7}$$

i.e. $Q$, $Q_A$ and $Q'$, $Q'_A$ form orthonormal pairs of functions.

Now consider the product, $Q'(x)A(x)Q(x)$. We have, from (I.6) and its analogue for P',

$$Q'(x)A(x)Q(x) = Q'(x)[\,Q(x) < A >_P + Q_A(x)\,\Delta_P A\,]$$

$$= [\,\Delta_{P'}A\,Q'_A(x) + < A >_{P'} Q'(x)\,]\,Q(x) . \tag{I.8}$$

Integrating through the equations and collecting expectation value terms on the left and rms deviation terms on the right, we have,

$$(< A >_{P'} - < A >_P)(Q', Q) = (Q', Q_A)\,\Delta_P A - \Delta_{P'}A\,(Q'_A{}^*, Q). \tag{I.9}$$

This yields the inequality,

$$|< A >_{P'} - < A >_P|\,(Q', Q) \le |(Q', Q_A)|\,\Delta_P A + \Delta_{P'}A\,|(Q'_A, Q)|. \tag{I.10}$$

The desired inequality would then follow from (I.3) and,

$$|(Q', Q_A)|,\ |(Q'_A, Q)| \le \sin\theta . \tag{I.11}$$

But (I.11) follows immediately from (I.3),

$$\|Q' - Q\,(Q, Q') - Q_A\,(Q_A, Q')\|^2 \ge 0 , \tag{I.12a}$$

and

$$\|Q - Q'\,(Q', Q) - Q'_A\,(Q'_A, Q)\|^2 \ge 0 , \tag{I.12b}$$

where, $\|F\|^2 := (F, F)$. □.

We will close this appendix with the remark that whenever marginal probability distributions obtained from P and P' can still yield probability distributions for the function, A, the inequality from the marginal distributions will never be weaker than that from the original P and P', in the sense that the angle θ from the marginals is less than or equal to the angle from P and P'. The strongest possible form of the inequality comes from using just the probability distributions for A itself.

**Appendix II:**

Define the non-negative quantity, d, by,

$$d := |<\Pi>_{\psi'} - <\Pi>_{\psi}|, \tag{II.1}$$

and the quantity, x, by,

$$x := <\Pi>_{\psi'} + <\Pi>_{\psi} - 1. \tag{II.2}$$

We then have, using (3.7b),

$$\frac{|<\Pi>_{\psi'} - <\Pi>_{\psi}|}{\Delta_{\psi'}\Pi + \Delta_{\psi}\Pi} = \frac{2\,d}{\sqrt{1-(x+d)^2} + \sqrt{1-(x-d)^2}} \quad , \tag{II.3}$$

where, for fixed d, the possible range of x is given by,

$$-1 + d \le x \le 1 - d \; . \tag{II.4}$$

At either end of that range, with the smaller expectation value of Π equal to zero or the larger expectation value equal to unity, (A.3) becomes,

$$\frac{|<\Pi>_{\psi'} - <\Pi>_{\psi}|}{\Delta_{\psi'}\Pi + \Delta_{\psi}\Pi} = \sqrt{\frac{d}{1-d}} \quad , \tag{II.5}$$

and, upon substitution into (3.7a), we obtain (3.8b), as asserted. For fixed d, (II.5) is the largest possible value of (II.3), and in the middle of the range for x, x = 0, (II.3) bottoms out at,

$$\frac{|<\Pi>_{\psi'} - <\Pi>_{\psi}|}{\Delta_{\psi'}\Pi + \Delta_{\psi}\Pi} = \frac{d}{\sqrt{1-d^2}} \ , \qquad \text{(II.6)}$$

which, when substituted into (3.7a) yields (3.8a) as providing the general upper limit for the absolute difference of projector expectation values.